\documentclass[structabstract]{aa}
\usepackage{textcomp}
\usepackage{natbib}
\bibpunct{(}{)}{;}{a}{}{,} 
\usepackage{graphicx}
\usepackage{txfonts}
\begin{document}
%
  \title{Oxygen depletion in dense molecular clouds: a clue to a low O$_2$ abundance?}

   \subtitle{ }

   \author{ U. Hincelin \inst{1,2}
          \and V. Wakelam \inst{1,2}  \and F. Hersant \inst{1,2}  \and S. Guilloteau \inst{1,2}   \and J.~C. Loison \inst{3,4} \and P. Honvault  \inst{5,6}
          \and J. Troe \inst{7,8}
                }

   \institute{Universit\'e de Bordeaux, Observatoire Aquitain des Sciences de l'Univers, 2 rue de l'Observatoire, BP 89, F-33271 Floirac Cedex, France
   			 \and CNRS, UMR 5804, Laboratoire d'Astrophysique de Bordeaux, 2 rue de l'Observatoire, BP 89, F-33271 Floirac Cedex, France 
         \and Universit\'e de Bordeaux, Institut des Sciences Mol\'eculaires, 351 Cours de la Lib\'eration, F-33405 Talence Cedex, France 
         \and CNRS UMR 5255, Institut des Sciences Mol\'eculaires, 351 Cours de la Lib\'eration, F-33405 Talence Cedex, France
         \and Universit\'e de Franche-Comt\'e, Institut UTINAM, UMR CNRS 6213, F-25030 Besan\c{c}on Cedex, France
         \and Laboratoire Interdisciplinaire Carnot de Bourgogne, UMR CNRS 5209, Universit\'e de Bourgogne, 9 av. A. Savary, F-21078 Dijon Cedex, France
         \and Georg-August-Universit$\rm \ddot{a}$t G$\rm \ddot{o}$ttingen, Institut f$\rm \ddot{u}$r Physikalische Chemie, Tammannstrasse 6, D-37077 G$\rm \ddot{o}$ttingen, Germany 
         \and Max-Planck-Institut $\rm f\ddot{u}$r Biophysikalische Chemie, Am Fassberg 11, D-37077 G$\rm \ddot{o}$ttingen, Germany
 }
   \date{Received xxx ; accepted xxx}


  \abstract
   {Dark cloud chemical models usually predict large amounts of O$_2$, often above observational limits. }
   {We investigate the reason for this discrepancy from a theoretical point of view, inspired by the studies of Jenkins and Whittet on oxygen depletion.}
   {We use the gas-grain code Nautilus with an up-to-date gas-phase network to study the sensitivity of the molecular oxygen abundance to the oxygen elemental abundance. We use the rate coefficient for the reaction O + OH at 10~K recommended by the KIDA (KInetic Database for Astrochemistry) experts. }
   {The updates of rate coefficients and branching ratios of the reactions of our gas-phase chemical network, especially N + CN and H$_3^+$ + O, have changed the model sensitivity to the oxygen elemental abundance. In addition, the gas-phase abundances calculated with our gas-grain model are less sensitive to the elemental C/O ratio than those computed with a pure gas-phase model. The grain surface chemistry plays the role of a buffer absorbing most of the extra carbon. Finally, to reproduce the low abundance of molecular oxygen observed in dark clouds at all times, we need an oxygen elemental abundance smaller than $1.6\times 10^{-4}$.}
   {The chemistry of molecular oxygen in dense clouds is quite sensitive to model parameters that are not necessarily well constrained. That O$_2$ abundance may be sensitive to nitrogen chemistry is an indication of the complexity of interstellar chemistry. }

   
   \maketitle
%

\section{Introduction}

Chemical models predict that gas-phase oxygen should be mainly in the form of O$_2$ and CO in the cold interstellar medium \citep[see for instance][]{2006A&A...451..551W,Quan2008}. Since the 1980's, there have been searches for O$_2$ in the interstellar medium using ground-based and space telescopes \citep[see][and references therein]{2003A&A...402L..77P}. First, analyses of data from the {\it SWAS} satellite gave an upper limit of about $10^{-6}$ in dense clouds \citep{2000ApJ...539L.123G}.
The ODIN satellite also initially gave negative results \citep{2003A&A...402L..77P} with improved upper limits of $\approx (1-2)\times 10^{-7}$ for nine sources. However, a reanalysis by \cite{2007A&A...466..999L}, using more precise knowledge of the telescope behavior, resulted in a detection of O$_2$ in $\rho$~Ophiuchi cloud, with a beam-averaged abundance of $5\times 10^{-8}$ relative to H$_2$. 
Using ground-based observations of $^{16}$O$^{18}$O and C$^{18}$O lines, \citet{2010A&A...510A..98L} argue that the emitting region may be much smaller than the beam of ODIN thus the O$_2$ abundance could be larger by one or two orders of magnitude. Regardless of the exact numbers, the sparsity of O$_2$ detections in the various target molecular clouds is an indication that this molecule may not be a reservoir of oxygen.

Many explanations have been proposed to reconcile observations and models.
From a chemical modeling point of view, pure gas-phase chemical models can explain the observed upper limits for clouds younger than $10^{5}$~yr \citep{2006A&A...451..551W}.  \citet{2001A&A...370..557V} explored the possibility that chemical models can display bistabilities \citep{1993ApJ...416L..87L}. In the parameter space where bistability exists, one of the solutions is characterized by a very low abundance of O$_2$. In this solution however, all molecular specie abundances are very small, including the CO abundance \citep{2006A&A...459..813W}, which is not what is observed. As another possibility, the effect of uncertainties in the rate coefficient of the main reaction of production of O$_2$ (O + OH $\longrightarrow$ O$_2$ + H) was explored by \citet{Quan2008}. One needs however to decrease this rate coefficient by a considerable amount to modify the predicted abundance of O$_2$.

In the presence of dust, molecular oxygen in the gas phase can be adsorbed onto grain surfaces.
Adsorption of O$_2$ onto dust grains is insufficient in itself to lower the O$_2$ abundance after $10^{6}$~yr, because of the balance with thermal evaporation and cosmic-ray-induced desorption. However, the adsorbed O$_2$ molecule can be successively hydrogenated to form HO$_2$ and H$_2$O$_2$. Then H$_2$O$_2$ reacts with H to form water \citep[see also][]{2002A&A...395..233R}, which is more difficult to release from the grains. This surface chemistry allows one to decrease the abundance of O$_2$ after $10^{6}$~years, but a peak in the abundance larger than observational limits remains between $10^{5}$ and $10^{6}$~years.
Unfortunately, this time range encompasses the \textquotedblleft age\textquotedblright \ of cold cores determined by the comparison between large network chemical models and observations of more than 30 species in TMC-1~(CP) and L134N~(N) \citep[see][]{2006A&A...451..551W,2007A&A...467.1103G,2004MNRAS.350..323S}.
\citet{2000ApJ...539L.129B} found a small abundance of O$_2$ in the gas at all times, in agreement with observations, using a simplified version of grain surface chemistry assuming conversion of O to H$_2$O and C to CH$_4$, and choosing the branching ratio of the reaction H$_3$O$^+$ + e$^-$ $\longrightarrow$ H + H$_2$O that equals 0.33 to reproduce the H$_2$O abundance. Experimental measurements from \citet{2000ApJ...543..764J} showed that this branching ratio is 0.25. Finally, \citet{2009ApJ...690.1497H} studied the chemistry of O$_2$ as a function of the depth in molecular clouds using a one-dimensional steady-state PDR model. The authors studied the influence of many parameters and found that the O$_2$ abundance peaks at Av between 4 and 6 and that molecules would be strongly depleted at larger Av. 



In this paper, we revisit the question of the O$_2$ abundance in dark clouds using the gas-grain model Nautilus with the most recent gas-phase network from the KIDA database and new insight into the oxygen elemental abundances provided by \citet{2009ApJ...700.1299J}. In the next two sections, we introduce the problem of the choice of oxygen elemental abundance for dense cloud chemical modeling and describe our chemical model. The results of our simulations and comparisons with observations in the two dark clouds L134N and TMC-1 are shown in section~\ref{results}. We present our conclusions about our work in the last section.

\section{Carbon and oxygen depletion}

The depletion of oxygen related to cosmic reference values has been investigated for many years. For instance, \citet{1998ApJ...493..222M} concluded that there was no evidence for density-dependent oxygen depletion from the gas phase based on the analysis of their observations of 13 stars using the {\it Goddard High Resolution Spectrograph (GHRS)} onboard the {\it Hubble Space Telescope (HST)}. They thus proposed a mean interstellar gas-phase oxygen abundance (O/H) of $3.19\times10^{-4}$.  A few years later, \citet{2004ApJ...613.1037C} found a weak correlation between the oxygen gas-phase abundance and the density of the clouds, analyzing a larger sample of sources observed with the {\it Space Telescope Imaging Spectrograph (STIS)} of {\it HST} and {\it Far Ultraviolet Spectroscopic Explorer (FUSE )}. At densities below 1~cm$^{-3}$, they found an O/H abundance of $3.9\times 10^{-4}$ and an abundance of $2.8\times 10^{-4}$ at densities above 1~cm$^{-3}$.

A comprehensive study of the depletion problem by \citet{2009ApJ...700.1299J} brought additional findings to light.
Jenkins re-analyzed archival data of atomic lines in more than two hundred lines of sight.
The main result of his analysis is that all elements (except nitrogen), and even sulphur, seem to deplete with density.
The depletion factor for carbon is much smaller than for oxygen (but mostly less certain) so that one expects an increase in the C/O gas-phase elemental ratio with the density of the cloud.
The main mechanism capable of explaining this depletion is accretion onto grains.
The depletion of oxygen, however, cannot be accounted for by the formation of silicates and oxides since the oxygen depletion appears to be larger than that of Mg + Si + Fe.
In the diffuse medium, densities are so low that the collision probability between grains and gas-phase species cannot explain the depletion. It is usually assumed that depletion occurs in denser regions and that the mass exchange between the dense and diffuse medium explains these observations \citep{1990ASPC...12..193D}. In a sense, the observation of the elemental depletion as a function of the cloud density traces the survival of refractory compounds to the UV radiation in the ISM. This may indicate that depletion of the elements is stronger in denser regions.
There is indirect evidence of this high depletion, such as the small abundance of SiO in dense clouds \citep{1989ApJ...343..201Z} compared to that of cosmic silicon.
To reproduce observations of gas-phase molecules in dense clouds, \citet{1982ApJS...48..321G} proposed to use a \textquotedblleft depletion factor\textquotedblright \ of ten on observed atomic abundances in diffuse clouds for all elements except He, C, N, and O. These elemental abundances constitute the largely used \textquotedblleft low metal" elemental abundances.
The value of the \textquotedblleft depletion factor" is however poorly constrained and remains a free parameter in chemical models. A sensitivity analysis by \citet{2010A&A...517A..21W} show that this is one of the most important parameters.

Carbon and oxygen are usually assumed not to display additional depletion compared to diffuse clouds but this view is now changing. Additional depletion of oxygen in an organic refractory  component of interstellar dust seems to be the most probable way of accounting for all oxygen in dense environments as shown by \citet{2010ApJ...710.1009W}. Following fresh insight from Jenkins about oxygen depletion, \citeauthor{2010ApJ...710.1009W} compiled an inventory of the different forms of O-bearing species as a function of cloud density from an observational point of view (see his Fig. 3). In diffuse clouds, oxygen is partly depleted in silicates and oxides (this is the fraction of cosmic oxygen already depleted in the diffuse gas), in atomic form (which is observed), and in an unidentified form (i.e. ``unidentified depleted oxygen" called UDO, possibly organic refractory compounds). As the density increases, the fraction of UDO increases whereas that of atomic oxygen decreases. At densities high enough for chemistry to be efficient (atomic oxygen then cannot be observed anymore), oxygen goes into gas-phase CO and ice compounds (mainly CO, CO$_2$ and H$_2$O). At hydrogen densities of around $1000$~cm$^{-3}$, 28\% of oxygen would be in gas-phase CO and icy species, 19\% would be in refractory silicate and oxide forms, and presumably 49\% would in UDO. The net result of this is that only 32\% of the oxygen would be available for the chemistry.
We note that the missing oxygen in the dust phase has been debated in \citet{2010A&A...517A..45V}.

\section{The model}\label{sec:model}

\subsection{Nautilus}\label{subsec:nautilus}

We used the Nautilus chemical model described in \citet{2009A&A...493L..49H}. The model solves the kinetic equations for gas-phase and grain surface chemistries. Details of the physical and chemical processes included in the model are given in a benchmark paper by \citet{2010A&A...522A..42S}. We use typical dense cloud conditions: a gas and dust temperature of 10~K, an H density of $2\times 10^4$~cm$^{-3}$, a visual extinction of 10, and a cosmic-ray ionization rate of $1.3\times 10^{-17}$~s$^{-1}$.  A single grain size of $\rm 0.1~\mu m$ is used to compute adsorption rates following \cite{1992ApJS...82..167H}. The cross-section per H nucleus used is  $5.72\times 10^{-22}~$cm$^{2}$ and we assumed thermal velocities.  The adsorption energy used for O$_2$ is 1000~K (Herma Cuppen private communication). As cosmic rays can penetrate deep into grains, we do not restrict the evaporation of species by cosmic rays to the surface layer of molecules.  The species are assumed to be initially in an atomic form as in diffuse clouds except for hydrogen, which is converted entirely into H$_2$.
Elements with an ionization potential below the maximum energy of ambient UV photons (13.6~eV, the ionization energy of H atoms) are initially in a singly ionized state, i.e., C, S, Si, Fe, Na, Mg, Cl, and P.

\subsection{Chemical network}

The chemical network,  adapted from \citet{2007A&A...467.1103G}, includes 6142 reactions, of which 4394 are pure gas-phase reactions and 1748 are grain-surface and gas-grain interaction reactions. The model follows the chemistry of 458 gas-phase species and 196 species on grains. The gas-phase network has been updated according to the recommendations from the experts of the KIDA database\footnote{http://kida.obs.u-bordeaux1.fr}. KIDA, for KInetic Database for Astrochemistry, is a recently opened online database of gas-phase reactions of interest for astrochemical (interstellar medium and planetary atmospheres) studies \citep[see][]{2010sf2a.conf..239W,2010SSRv..156...13W}. Recommendations on rate coefficients by experts in physico-chemistry are given for key reactions.

The rate coefficient of the reaction O + OH $\longrightarrow$ O$_2$ + H is a subject of debate \citep[see][]{Quan2008}. The KIDA experts suggest that the rate coefficient at 10~K is between $2\times10^{-11}$ and $\rm 8\times10^{-11}~cm^{3}~molecule^{-1}~s^{-1}$  (see section~\ref{subsubsec:O+OH}).
In the rest of the paper, we use the lower limit for this rate coefficient. Results using the upper limit are discussed in section~\ref{sec:rates}.

 \subsubsection{Experimental and theoretical determination of the rate coefficient of the reaction O + OH $\longrightarrow$ O$_2$ + H} \label{subsubsec:O+OH}

The study of this reaction has attracted considerable experimental attention \citep{1980CPL....69...40H,1981JCSFT2_77_997,1980JPhCh_84_3495,1983JPhCh_87_4503,1994JCSFT_90_3221,2002CPL_358_157-62,2006JPCA_110_6673,2006JPCA_110_3101}, and there have also been a large number of theoretical studies using a variety of methods \citep{2000JChPh.11311019H,2001JChPh.115.3621T,2007JChPh.127b4304X,2008JChPh.128a4303L,2009JChPh.131v1104L,2009PhRvA..79b2703Q,2004NAS_21-44,2009JChPh.130r4301J,2010JChPh.133n4306L}. The experimental rate constant is well determined between 140 and 300~K decreasing from $7\times10^{-11}$~cm$^{3}$~molecule$^{-1}$~s$^{-1}$ at 140~K to $3\times 10^{-11}$~cm$^{3}$~molecule$^{-1}$~s$^{-1}$ at 300~K. Between 40 and 140~K, the reaction has been studied in a CRESU (Cin{\'e}tique de R{\'e}action en Ecoulement Supersonique Uniforme) apparatus \citep{2006JPCA_110_3101} leading to a value of around $\rm 3.5(\pm1.0)\times10^{-11}~cm^{3}~molecule^{-1}~s^{-1}$ which however has large uncertainties. Quasi-classical trajectory calculations give good agreement with experiment between 300~K and 3000~K \citep{2001JChPh.115.3621T} and between 40~K and 140~K \citep{2008CPL...462...53J} but the relatively good agreement at low temperature may be fortuitous. The reaction, which proceeds through a relatively long-lived HO$_2$ complex, should be amenable to a statistical treatment and the statistical adiabatic channel model should be appropriate \citep{2000JChPh.11311019H,2001JChPh.115.3621T} but has to deal with dynamical barriers.
Surface hopping forward and backward between adiabatic channel potentials on several electronic states allows the system to avoid the dynamical bottleneck to some extent leading to a marked increase in the rate constant around 50~K.
The rate constant calculated by this method \citep{2000JChPh.11311019H} may be considered as an upper value.
Time-dependent wave packet methods are unsuitable for the low temperature regime leading to an unreliably low rate that is constant at low temperature \citep{2007JChPh.127b4304X,2008JChPh.128a4303L}.
Time-independent quantum mechanical calculations, supposed to be the more accurate at low temperature, have been applied to this system leading to value around $\rm 4\times10^{-11}~cm^{3}~molecule^{-1}~s^{-1}$ at 10~K \citep{2009JChPh.131v1104L,2009PhRvA..79b2703Q}.
However, they neglect spin orbit coupling and electronic fine structure of O and OH, as well as surface hopping dynamics between the ground and excited potentials at large O-OH distances.
Among these effects, surface hopping dynamics is suspected to be important at low temperature \citep{2004NAS_21-44,2010JChPh.133n4306L} and the rate constant may be as high as $\rm 8\times10^{-11}~cm^{3}~molecule^{-1}~s^{-1}$ at 10~K.
Taking into account the various calculations and measurements, the KIDA experts recommend that the rate coefficient at 10~K is between $2\times10^{-11}$ and $\rm 8\times10^{-11}~cm^{3}~molecule^{-1}~s^{-1}$.

\subsection{Elemental abundances}\label{elements}

\begin{table}
\caption{Elemental abundances (/H)}
\begin{center}
\begin{tabular}{lc|lc}
\hline
\hline
Element & Abundance & Element & Abundance \\
\hline
He & 0.09$^a$ & N & 6.2(-5)$^b$ \\
C & 1.7(-4)$^b$ & O & (3.3-2.8-1.8-1.4)(-4)$^c$ \\
S & 8(-8)$^d$ & Si & 8(-9)$^d$ \\
Fe & 3(-9)$^d$ & Na & 2(-9)$^d$ \\
Mg & 7(-9)$^d$ & Cl & 1(-9)$^d$ \\
P & 2(-10)$^d$ &  & \\
\hline
\end{tabular}
\end{center}
\tablefoot{
$^a$ see \citet{2008ApJ...680..371W}
$^b$ from \citet{2009ApJ...700.1299J}.
$^c$ see text
$^d$ Low Metal elemental abundances
}
\label{elem_ab}
\end{table}

For each element, we define the elemental abundance as the ratio of the number of nuclei both in the gas and on the dust grains to the total number of H nuclei.
This excludes the nuclei locked in the refractory part of the grains. With this definition, the elemental abundance would be the \textquotedblleft gas phase elemental abundance\textquotedblright \ at high temperature, when all ices are sublimated.
Since the amount of the oxygen depletion is only based on indirect measurements, we consider four values for the oxygen elemental abundance :
1) $3.3\times 10^{-4}$: a ``low depletion'' case, using the gas-phase abundance observed in the diffuse cloud $\zeta$\,Oph, frequently used as a reference derived by \citet{2009ApJ...700.1299J} (this is also the mean abundance observed by \citet{1998ApJ...493..222M}); 2) $2.4\times 10^{-4}$:  an ``intermediate depletion''; 3) $1.8\times 10^{-4}$, following \citet{2010ApJ...710.1009W}: this value is very close to the ``low metal'' case; 4) $1.4\times 10^{-4}$: a ``high depletion'' case determined from Fig.16 of \citet{2009ApJ...700.1299J} with an extrapolation to a density of $2\times10^{4}$~cm$^{-3}$.


For simplicity, we use the ``low metal'' elemental abundances for the other elements except for He, C, and N. Carbon is in a similar situation as oxygen in the sense that no additional depletion is usually assumed, although \citet{2009ApJ...700.1299J} showed that it does exist the trend being much weaker and less robust. From the observations, \citet{2009ApJ...700.1299J} determined an atomic carbon abundance of $1.7\times 10^{-4}$ in $\zeta$\,Oph. We use this value, as the extrapolation to a density of $2\times 10^{4}~$cm$^{-3}$ using Jenkins' relations would only  reduce the carbon abundance to $1.2\times 10^{-4 }$. 
The C/O elemental ratio corresponding to our four models are 0.5, 0.7, 0.9, and 1.2. Nitrogen is the only element that was not found by \citet{2009ApJ...700.1299J} to deplete with density. Although it may be an observational bias as argued by Jenkins, we use the gas-phase abundance observed in $\zeta$\,Oph of $6.2\times 10^{-5}$. The helium abundance is assumed to be 0.09 \citep[see][for discussion]{2008ApJ...680..371W}. All elemental abundances are listed in Table~\ref{elem_ab}. 

\section{Results of the chemical simulations}\label{results}

\subsection{Computed abundances}

\begin{figure}
	\centering
		\includegraphics[width=0.5\textwidth]{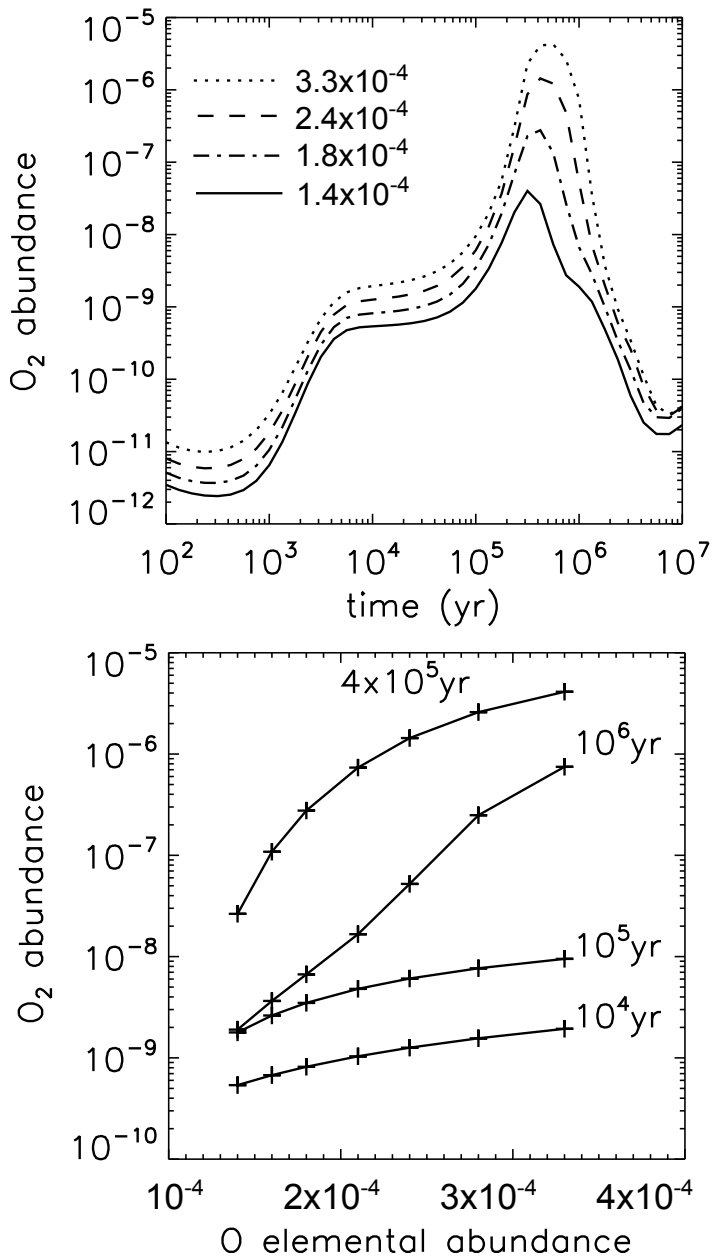}
	\caption{O$_2$ abundance (/H) as a function of time for four values of oxygen depletion (top) and as a function of oxygen elemental abundance for four ages (bottom).}
	\label{fig:O2ab_VS_time_and_O_ElemAb}
\end{figure}

Figure \ref{fig:O2ab_VS_time_and_O_ElemAb} shows the computed abundances of $\rm O_2$ in the gas phase as a function of time for the four elemental abundances of oxygen described in section~\ref{elements} (left panel) and as a function of the oxygen elemental abundance for different times (right panel).
A decrease in the oxygen elemental abundance produces a general decrease in the $\rm O_2$ abundances at any time, although the effect is stronger between $3\times 10^5$ and $2\times 10^6$~yr. Assuming that O$_2$ has been searched for in a variety of clouds with ages across this range, the non-detection of O$_2$ with abundances above $10^{-7}$ compared to  total hydrogen would require the elemental abundance of oxygen to be smaller than $1.6\times 10^{-4}$.


\begin{figure}
	\centering
		\includegraphics[width=0.5\textwidth]{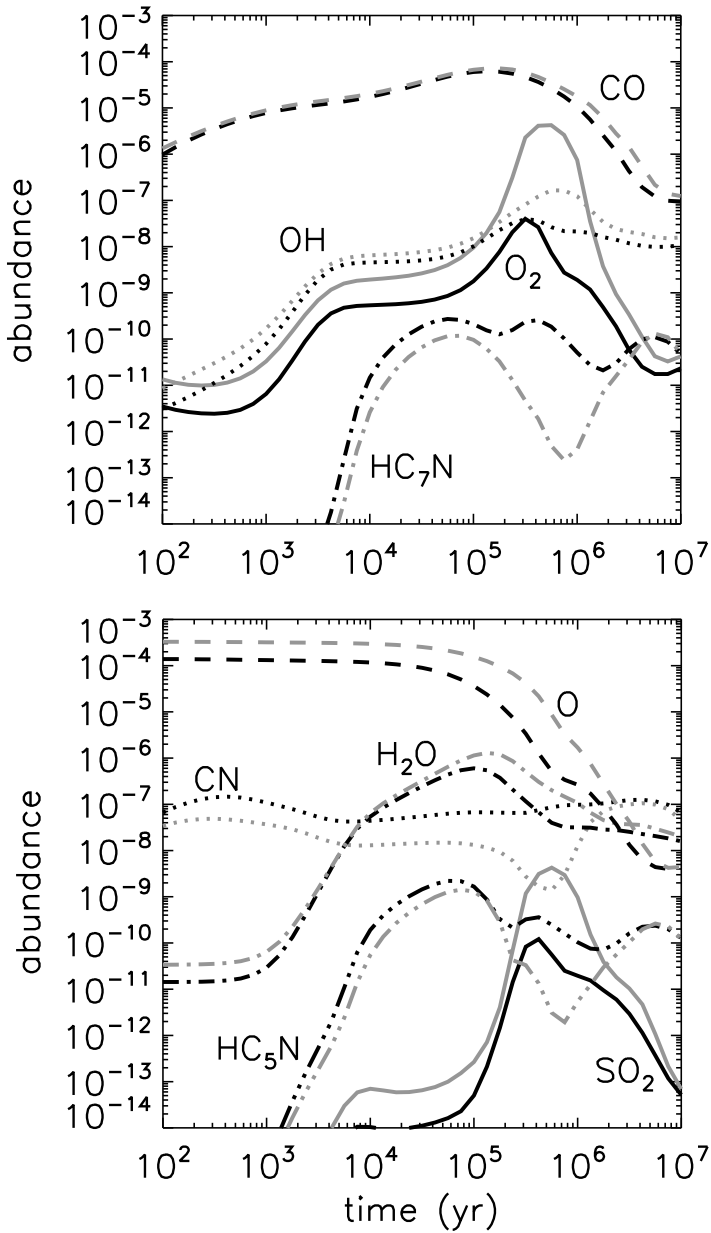}
	\caption{Gas-phase abundances relative to total hydrogen of a selection of species as a function of time computed for two different oxygen elemental abundances: $3.3\times10^{-4}$ (grey line) and $1.4\times10^{-4}$ (black line).}
	\label{fig:paper1_AbX_vs_time}
\end{figure}

Figure \ref{fig:paper1_AbX_vs_time} shows the computed abundances of a selection of gas phase species as a function of time, for the extreme "low" and "high depletion" cases. As expected, the abundances of carbon-rich species (such as cyanopolyynes) are higher in the "high oxygen depletion" case (higher C/O ratio), whereas the abundance of O-bearing species is lower. The various O-bearing species are however not influenced to the same extent. CO, OH, and H$_2$O are changed only slightly.
O$_2$ and SO$_2$ are lower by two orders of magnitude and more than one order of magnitude, respectively, at the peak abundance ($\sim 4\times 10^5$~yr).

In our four cases, the elemental C/O ratio varies over a large range. Pure gas-phase chemical models would be very sensitive to these variations: for example, \citet{2010A&A...517A..21W} show that the HC$_7$N abundance can be modified by four orders of magnitude at $10^7$~yr when C/O goes from 0.7 to 1. In our present study  and at $10^{7}$~yr, the HC$_7$N abundance does not depend much on  the different values of C/O that we adopted, as can be seen in Fig.~\ref{fig:paper1_AbX_vs_time}. When the C/O ratio increases, this occurs because the available C is mainly used to form C-rich molecules, e.g. C$_n$H$_m$, on the grains \citep[see also][]{2007A&A...467.1103G}. However, the modeled abundances are not only sensitive to the elemental C/O ratio but also to the elemental abundances themselves. Increasing both elemental abundances by a factor of two would for instance increase the CO gas-phase abundance by two orders of magnitude at $5\times 10^6$~yr in all the models, but the O$_2$ abundance at the peak ($\sim 4\times 10^5$~yr) remains unchanged. In this dense cloud modeling, the CO abundance in the gas phase decreases after a few $10^5$ yr because CO sticks onto grains and is then hydrogenated to form H$_2$CO and CH$_3$OH. When the C and O elemental abundances are increased, the gas phase CO abundance, and as a consequence the solid CO abundance, increases and takes longer to decrease its abundance. If we allow the system to evolve up to $10^8$ yr, the CO gas phase abundance is only two times larger than the one obtained with our previous elemental abundances. Decreasing C and O elemental abundance by a factor of two would slightly decrease the O$_2$ abundance.


 We started our chemistry assuming that all species were initially atomic, except for hydrogen.
If we instead assume that all carbon is initially in carbon monoxide, the results change drastically before $10^{5}$~years.
Carbon chains are obviously the most affected species.
Using these initial abundances increases the time taken to form the carbon-bearing molecules observed in dense clouds by a factor of between $10^{2}$ and $10^{6}$ depending on the molecule.

\subsection{Effects of the new rate coefficients}\label{sec:rates}

Among the updates of the network, the new values for the rate coefficients and branching ratios of the reactions O + C$_n$H  ($n$=2,3), O + C$_2$, C$^+$ + H$_2$ and particularly O + H$_3^+$ \citep[see also][]{2010SSRv..156...13W} and N + CN have changed the model sensitivity to the oxygen elemental abundance. In the ``high depletion'' case, the new rates result in smaller O$_2$ abundances ($4\times 10^{-8}$ instead of  $3.3\times 10^{-7}$ at the abundance peak near $\sim 4\times 10^5$~yr), while this abundance is not affected in the ``low depletion'' case. The O$_2$ abundance is also unexpectedly sensitive to the reaction N + CN. Decreasing the N + CN rate coefficient by a factor 3 increases the CN abundance at $3\times 10^{5}$~yr by a factor of $2.3$, and the reaction CN + O$_2$ becomes an efficient destruction channel reducing the O$_2$ abundance  (see also discussion about CN + O$_2$ reaction in section~\ref{sec:agreement}). This shows that the chemistry of a relatively simple molecule can be difficult to predict, and that unexpected reactions can be important. These results underline the importance
of using accurate rate coefficients in all circumstances.
Note that using the upper limit instead of the lower one for O + OH rate coefficient increases the abundance peak of O$_2$ by a factor two in the high depletion case, and the maximum elemental abundance of oxygen required to reproduce O$_2$ abundance in dark cloud is changed to $1.5\times 10^{-4}$.


\subsection{Agreement with observations in dark clouds}\label{sec:agreement}


\begin{table}
\caption{Observational values of abundances relative to total hydrogen in TMC-1(CP) and L134N (N) for some species.}
\begin{center}
\begin{tabular}{lcc}
\hline
\hline
Species & n(i)/nH$^{a}$ observed & n(i)/nH$^{a}$ observed \\
 & (TMC-1 (CP)) & (L134N (N)) \\
\hline
O$_2$ & $\leq$3.9(-8)$^b$ & $\leq$8.5(-8)$^b$ \\
CO & 4(-5)$^e$ & 4(-5)$^e$ \\
OH & 1(-7)$^c$ & 3.8(-8)$^e$ \\
H$_2$O & $\leq$3.5(-8)$^d$ & $\leq$1.5(-7)$^d$ \\
SO$_2$ & $<$5(-10)$^e$ & $\leq$8(-10)$^f$ \\
CN & 2.5(-9)$^c$ & 4.1(-10)$^e$ \\
HC$_5$N & 2(-9)$^g$ & 5(-11)$^e$ \\
HC$_7$N & 5(-10)$^g$ & 1.0(-11)$^e$ \\
\hline
\end{tabular}
\end{center}
\tablefoot{
$^a$ $a(b)=a\times 10^b$;
$^b$ from \citet{2003A&A...402L..77P};
$^c$ see \citet{2004MNRAS.350..323S};
$^d$ from \citet{2000ApJ...539L.101S};
$^e$ from \citet{1992IAUS..150..171O};
$^f$ from \citet{2000ApJ...542..870D};
$^g$ from \citet{1998cpmg.conf..205O}
}
\label{obs_ab}
\end{table}

We compared the abundances predicted by our model for different depletion cases ("high", "intermediate" and "low") with observations of gas-phase molecules in two cold clouds TMC-1 (CP) (the so-called cyanopolyyne peak) and L134N (N) (north peak) to check for the effects of oxygen depletion.
The observed abundances of some relevant species are listed in Table~\ref{obs_ab}. A more complete list can be found in \citet{2007A&A...467.1103G}.
The observational limits on the O$_2$ abundance relative to total hydrogen in TMC-1 and L134N are $3.85\times 10^{-8}$ and $8.5\times 10^{-8}$  \citep[see][]{2003A&A...402L..77P}, respectively, and they are reproduced by our models.
SO$_2$ is less efficiently produced in the gas-phase, which is in closer agreement with the observations, whereas CN is now overproduced  (CN abundance relative to total hydrogen in TMC-1~(CP) and L134N~(N) are, respectively, $2.5\times 10^{-9}$ \citep[see][]{2004MNRAS.350..323S} and $4.1\times 10^{-10}$ \citep[see][]{1992IAUS..150..171O}). Because of the importance of CN to the destruction of O$_2$ underlined in section~\ref{sec:rates}, this excess of CN may artificially destroy O$_2$. However, reducing the rate coefficient of CN + O$_2$ by a factor of 10 (to simulate a lower abundance of CN) does not affect the O$_2$ abundance, because other destruction channels take over, in particular C + O$_2$ $\longrightarrow$ CO + O.
Raising the rate coefficient of CN + O$_2$ would however continue to lower O$_2$ abundance. 

\begin{figure}
	\centering
		\includegraphics[width=0.5\textwidth]{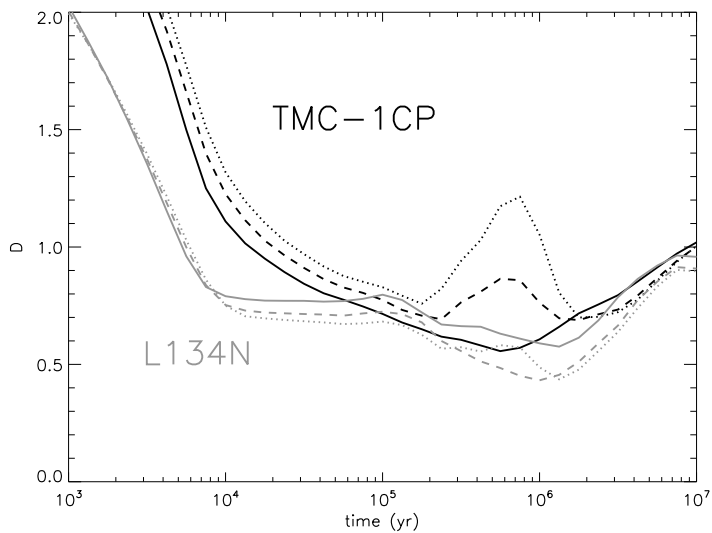}
	\caption{Parameter D as a function of time for TMC-1 (black lines) and L134N (grey lines) and three different oxygen elemental abundances: $3.3\times 10^{-4}$ (dotted line), $2.4\times 10^{-4}$ (dashed line), and $1.4\times 10^{-4}$ (solid line).}
	\label{fig:paper1_obsTMC1_L134N_vs_model}
\end{figure}

For a more global view, we define a pseudo-distance D between the models and the observations
\begin{equation}
\textit{D(t)}=\frac{1}{N}\sum_j\left|\log\left(X_{j}^{mod}\left(t\right)\right)-\log\left(X^{obs}_j\right)\right|,
\end{equation}
where $X^{obs}_j$ is the observed abundance of species j and $X^{mod}_j\left(t\right)$ the abundance of species j computed by the model at the time $t$, and N is the number of observed species, i.e., 53 in TMC-1 and 42 in L134N. The smaller the value of D, the closer the agreement. D is displayed for our three main models in Fig.~\ref{fig:paper1_obsTMC1_L134N_vs_model}. The O$_2$ abundance predicted by our model for the ``low'', ``intermediate'' and ``high'' depletion cases at the time of maximum agreement are, respectively, $4\times 10^{-9}$, $3\times 10^{-9}$, and $8\times 10^{-9}$ for TMC-1 and $3\times 10^{-8}$, $7\times 10^{-8}$, and $10^{-9}$ for L134N.
The elemental O abundance does not significantly affect the agreement in the case of L134N. However, as expected, the agreement in TMC-1 is closer in the ``high'' oxygen depletion case between $2\times10^{5}$ and $2\times10^{6}$ years.  This closer agreement comes from the higher abundance of cyanopolyynes during this range of time.

 We must keep in mind that the data obtained by observations indicate that the average abundance along the line of sight, and the abundances of species varies very considerably from the cloud surface to the cloud interior where the photodesorption is less important, as pointed out by \cite{2009ApJ...690.1497H}.




\section{Conclusions}

Our study has demonstrated that the low abundance of O$_2$ in dark clouds can be explained by gas-grain chemical models based on the fresh insight into elemental oxygen depletion obtained by \citet{2009ApJ...700.1299J} and \citet{2010ApJ...710.1009W}. High oxygen depletion also improves the overall agreement between models and observations for other molecules.
We have found that gas-grain models are less critically sensitive to the C/O ratio than pure gas chemistry. This is fortunate because the elemental C depletion remains poorly constrained by current observational studies. This limited sensitivity makes our conclusion more robust.

Besides this important result about the impact of the O depletion on the O$_2$ abundance, our study has also revealed that unexpected reactions may become very significant when the elemental abundances are modified. The example of the N + CN reaction is notable in this respect.  This may have consequences for models using limited reaction networks: while they may be valid over a limited range of input parameters (and for a limited number of predicted species), they should not be utilized in other conditions without re-assessing their performance.

\begin{acknowledgements}
This research was partially funded by the program PCMI from CNRS/INSU. U.H. is funded by a grant from the french ``R{\'e}gion Aquitaine''. The authors thank Anne Dutrey for helpful discussions.  We are grateful to Marcelino Ag{\'u}ndez for helpful comments on the manuscript. The authors also thank the referee for making this paper clearer and more useful to the reader.
\end{acknowledgements}

\bibliographystyle{aa}

\bibliography{aamnem99,biblio}

\end{document}